\documentstyle[11pt]{article}

\large
\begin{document}
\setlength{\baselineskip}{27pt}
\pagestyle{empty}
\vfill

\eject

{\small SUNY-NTG-94-10}
\vskip 0.5cm
\renewcommand{\thefootnote}{\fnsymbol{footnote}}
\centerline{\bf\LARGE Two-loop Correction to the Instanton Density}
\centerline{\bf\LARGE for the Double Well Potential}
\vskip 1cm
\centerline{\bf Christian F. W\"{o}hler and Edward Shuryak}
{\it Physics Department,
State University of New York, Stony Brook, NY 11794}
\vskip 1cm
\centerline{\bf Abstract}
   Feynman diagrams  in
 the instanton background are used
 for the calculation of  the tunneling amplitude,
up to the two-loops order.  Some  mistakes made in the previous works
are corrected. The same method is applied to
the next-order corrections to the ground state wave
function.\\

   A  motion in the double-well potential is a
classic example, displaying tunneling phenomena in their simplest form.
its discussion was revived
with the development of semiclassical methods in
quantum field theory, related to Euclidean classical solutions, the {\it
instantons} \cite{Belavin_etal,tHooft},
  (see e.g. good review
 \cite{abc of inst}).

   Although many questions concerning ensemble of instantons in QCD
 remain unclear,
  it was recently demonstrated that
 the instanton-induced effects are more important
than previously believed, dominating even such
quantities as baryonic masses, splittings and wave functions \cite{recent}.
Evaluation of the next-order corrections,
 going beyond the standard semiclassical
theory, are of great interest. As it is technically difficult to do such
calculations for the gauge theory, one may naturally
 start with the quantum mechanical example.
Note also, that some of the higher-dimensional problems
can be directly connected
to the double-well considered here.

   This problem was first addressed in ref. \cite{Alejnik}. Starting with
the anharmonic oscillator, it was shown how to apply Feynman diagrams
in quantum mechanics for a few simple problems\footnote{This part
 can be recommended as methodical examples in quantum field
theory courses: here Feynman diagrams appear without singularities of
divergences, and the results can be easily checked by standard means.},
and then the two-loop diagrams in the instanton background were evaluated.
The nontrivial point here is the new type of diagram,
coming from Jacobian of the
collective mode.

  Later an error in the Green's function used
 was noticed by \u{S}.Olejnik \cite{Olejnik}, who corrected it and also made
a new calculation of the instanton density, However, his
results strongly  disagree not only
with those in \cite{Alejnik}, but also with numerical simulations
 of path integrals  \cite{Shuryak} and (which was discovered later)
with the {\it exact}
result\footnote{Zinn-Justin work is based on solution of Schroedinger
equation in  a form of series expansion: thus it certainly cannot be
generalized
to the field theory context we are mainly interested in.
} obtained by  Zinn-Justin \cite{Zinn-Justin}.
One may suspect either some fundamental problem
        of the method used; or a simple error. As we show in
this letter, the method is fine and the error is found.

   Let us introduce our notations (same as in \cite{abc of inst}). A particle
of mass m=1 moves in the potential
\begin{equation}\label{dwpot}
V_{inst} = \lambda (x^{2}-\eta^{2})^{2}
\end{equation}
The instanton solution is
$X_{inst}(t) = \eta tanh(\frac{1}{2} \omega (t-t_{c}))$
where $t_{c}$ is the instanton center and
$\omega^{2} = 8 \lambda \eta^{2}$.
Setting $\omega = 1$,
and shifting one minimum to the origin
one gets the anharmonic oscillator potential
$ V  = \frac{1}{2} x^{2} - \sqrt{2\lambda} x^{3} + \lambda x^{4} $
with one small dimensionless parameter $\lambda$, to be used for the expansion.
The classical action of the instanton  solution
$S_{0}  = \frac{1}{12 \lambda}$ is therefore  large, and we are looking for the
corrections to the wave functions of the first and second state (i=1,2), the
ground state energy and the splitting
in the following form
\begin{eqnarray}
\Psi_{i}(\eta) & = & \Psi_{i}^{0}(\eta) (1 + A/S_{0} + \cdots) \\
E_{0} & = &E_{0}^{0}(1 + B/S_{0} + \cdots) \\
\delta E & = & \Delta E(1 + C/S_{0} + \cdots)
\end{eqnarray}
Here $E_{0}^{0} = \frac{\omega}{2}$ and the (one-loop) semiclassical result for
the splitting is
$\Delta E = 2 \omega \sqrt{6 S_{0}/\pi} e^{-S_{0}}$.
Up to exponentially small corrections $\Psi_{0}(\pm\eta)$ and
$\Psi_{1}(\pm\eta)$ are the symmetric and antisymmetric linear
combinations of the ground state wavefunction of the perturbed harmonic
oscillator, and therefore $A_1=A_2=A$.

 The transition amplitude
for $1/\Delta E >> \tau >>1 $ can therefore be written as
\begin{eqnarray}
\langle -\eta \mid e^{-H\tau} \mid \eta \rangle \longrightarrow
\mid \Psi^{0}_{0}(\eta) \mid ^{2}
(1 + 2A/S_{0} + \cdots) \times \nonumber \\
 \times e^{-\omega/2 (1 + B/S_{0} + \cdots) \tau}
\Delta E (1 + C/S_{0} +\cdots) \tau
\end{eqnarray}
 Normalizing it to the transition amplitude
$\langle \eta \mid exp(-H\tau) \mid \eta \rangle$ for
the corresponding
anharmonic oscillator,
one can get rid of unwanted corrections A and B.
In the Feynman path integral language,  this ratio can be
 calculated by computing the diagrams in the instanton background ( with
the instanton-based vertices and the instanton-based Green's function ), and
then by subtracting the diagrams for
 the anharmonic oscillator (see \cite{Alejnik} for
details).

 The instanton-based  Green's
function should satisfy the standard
 differential equation
and also be orthogonal to the zero mode.
The result \cite{Olejnik} is
\begin{equation}
G(x,y) = G_{0}(x,y)[2-xy+1/4 \mid x-y \mid (11-3xy) + (x-y)^2]
	 +3/8(1-x^2)(1-y^2)[\ln(G_{0}(x,y)) -11/3]
\end{equation}
where $x =  \tanh(t/2),
y  =  \tanh(t'/2)$ and

\begin{eqnarray}
G_{0}(x,y)  =  \frac{1- \mid x-y \mid - xy}{1 + \mid x-y \mid - xy}
\end{eqnarray}

 The correction C we are interested in can be written as
four diagrams for the instanton
correction \cite{Alejnik}
$C = a_{1} + b_{11} + b_{12} + c_{1}$ as
shown in the
Figure. Three first terms
 appear also for the anharmonic oscillator and should be subtracted,
the last one comes from Jacobian of the zero mode and has no
analogs in this case.
Their analytic expressions are

\begin{eqnarray}
a_{1} & = & -3 \beta \int^{\infty}_{-\infty}dt (G^{2}(t,t) - G^{2}_{0}(t,t))
\nonumber \\
      & = & -\frac{97}{1680} \\
b_{11} & = & 3 \alpha^{2} \int^{\infty}_{-\infty} \int^{\infty}_{-\infty}dt dt'
(\tanh(t/2)\tanh(t'/2)G^{3}(t,t') - G^{3}_{0}(t,t')) \nonumber \\
     & = & -\frac{53}{1260} \\
b_{12} & = & \frac{9}{2} \alpha^{2} \int^{\infty}_{-\infty}
 \int^{\infty}_{-\infty}
dt dt'(\tanh(t/2) \tanh(t'/2)G(t,t)G(t,t')G(t',t') \nonumber \\
     &   &- G_{0}(t,t)G_{0}(t,t')G_{0}(t',t')) \nonumber \\
     & = &-39/560 \\
c_{1} & = & -9 \beta \int^{\infty}_{-\infty} \int^{\infty}_{-\infty}dt dt'
\frac{\tanh(t/2)}{cosh^{2}(t/2)}\tanh(t'/2)G(t,t')G(t',t') \nonumber \\
    & = & -49/60
\label{diagrams}
\end{eqnarray}
from which follows $C = -71/72$.

This is exactly the result of J. Zinn-Justin in \cite{Zinn-Justin}
\footnote{The error made in ref.\cite{Olejnik} was found to be related with
diagrams $b_{11},b_{12}$, in which the time-dependent triple vertex
 those were erroneously
included for the subtracted anharmonic oscillator part as well.}
{}.
For comparison,
 Alejnikov and Shuryak \cite{Alejnik} and Shuryak \cite{Shuryak}
got $C\approx -1$, and Olejnik \cite{Olejnik} had $C\approx 0$.

  The findings  that (i) the correction has a coefficient C=O(1) and
(ii) it is {\it
negative} are important: they demonstrate that (i) the
usual (one-loop) semiclassical
 formula does become inaccurate for not-so-large barriers ($S\sim 1$), and
(ii) that it {\it overestimates} the tunneling probability.
 The magnitude of the
corresponding coefficient in gauge theories remains unknown, although
this question is of great practical importance.

   Let us now switch to our second subject,
the wave function correction (denoted A in (5)).
Although it is much simpler than the problem discussed above
(and has no dramatic history), it has a potential  trap, which
was not discussed for our knowledge and is worth
mentioning.

Now, one has  the general anharmonic potential
$V_{anh osc}(x) = x^{2}/2 + \alpha x^{3} + \beta x^{4}$
and evaluates the transition amplitude $\langle x=0 \mid e^{-H_{anh osc}
\tau} \mid x=0 \rangle$. For $\tau$ going to infinity it
gives terms proportional
to $\tau$ which are the energy corrections and constant terms which are the
wavefunction corrections.

Two different Green's functions can be used,
which satisfy both the same differential equation but vanish at different
boundaries.
The infinite boundary Green's function is simply
$G_{infinite}(t,t') = 1/2 e^{-\mid t-t' \mid}$
while the finite one is
$$G_{finite}(t,t') = \sinh(\tau/2-t)\sinh(t'+\tau/2)/\sinh(\tau)
\,\,\,\,\,(t \geq t') $$
\begin{equation}
= \sinh(\tau/2-t')\sinh(t+\tau/2)/\sinh(\tau) \,\,\,\,\,(t' \geq t)
\end{equation}

  Both give the same result for the energy shift, but {\it different} ones
for the wave function correction. Up to
the first
order in $\beta$ and $\alpha ^{2}$
\begin{equation}
A_{infinite} = -(29/96) \alpha^{2}; \, \,\,
 A_{finite} = (9/16) \beta -(167/96) \alpha ^{2}
\end{equation}
Comparison of the two results to the corrections obtained by old-fashioned
perturbation theory
shows that  $A_{finite}$ is correct\footnote{
Note that this observation does not invalidate the calculations above
for the constant C:
because the wave function corrections are the same for both
the instanton and the anharmonic oscillator, up to exponentially small terms
 the constants A,B are cancelled whatever Green's function is used.
}.
  The lesson is as follows: unlike the
energy correction (which corresponds to infinitely long paths), the wave
function is a kind of a boundary effect, and therefore
proper restrictions for the quantum paths (and Green's functions)
should not be ignored.

   In conclusion, we have calculated the tunneling amplitude (equal to the
energy splitting or the
instanton density) up to two-loops, using  Feynman diagrams.
Two important elements of the calculation, making it somewhat less
standard, is the {\it non-trivial background field} and {\it
new effective vertices},
which come the zero-mode Jacobian rather than from the action.
Our results agree
exactly  with ref.
\cite{Zinn-Justin}, based on a different method, and
with other works
inside the accuracy. In ref.\cite{Olejnik} the error is found.

  Extensions of this work can be (i) three and more
 loop corrections
for quantum mechanical examples, and,
 more important (ii) possible two-loop calculations for gauge
theories, in
essentially the same setting.
Such calculation would be of great interest, because
the problem of large-size (or small action) instantons in QCD remains
unsolved.

\newpage
\begin{picture}(350,400)
\thicklines
\put(50,320){\circle{50}}
\put(50,280){\circle{50}}
\put(0,296){$(-3 \beta)$}
\put(45,250){$a_{1}$}
\put (140,300){\line(1,0){40}}
\put (160,300){\circle{50}}
\put(105,296){$(3 \alpha^{2})$}
\put(160,250){$b_{11}$}
\put(280,300){\circle{20}}
\put(290,300){\line(1,0){30}}
\put(330,300){\circle{20}}
\put(235,296){$(\frac{9}{2} \alpha^{2})$}
\put(300,250){$b_{12}$}
\put(0,146){$\ddot{X}_{inst}$}
\put(40,150){\circle*{5}}
\put(40,150){\line(1,0){30}}
\put(80,150){\circle{20}}
\put(45,100){$c_{1}$}
\put(60,160){$\alpha$}
\put(0,0){\makebox(350,50){Fig. 1.} }
\end{picture}
\vskip 1cm
\centerline{\bf Caption}
  Feynman diagrams for the two-loop
 correction to the energy splitting C. Here
  the triple and
quartic coupling constants are $\alpha=-\sqrt{2\lambda}+4X_{inst}(t),
\beta=\lambda$ for the instanton and $\alpha=-\sqrt{2\lambda},
\beta=\lambda$ for subtracted anharmonic oscillator.
  The indicated numbers of combinatorial
origin should be added to ordinary Feynman rules,   $X_{inst}$ in the last
diagram stands for the classical instanton path, dots are time derivatives.
\end{document}